# Exploring the Temporal Variation of the Solar Quadrupole Moment $J_2$

Saliha Eren[1] and Jean-Pierre Rozelot[2]
[1] The Arctic University of Norway, Department of Physics and Technology, Postboks 6050 Langnes, NO-9037 Tromsø, Norway; saliha.eren@uit.no
[2] Université Côte d'Azur, 77 Chemin des Basses Moulières, F-06130 Grasse, France; jean-pierre.rozelot@grenoble-inp.org

## Abstract

Recently, Rozelot & Eren pointed out that the first solar gravitational moment ($J_2$) might exhibit a temporal variation. The suggested explanation is through the temporal variation of the solar rotation with latitude. This issue is deeper developed due to an accurate knowledge of the long-term variations in solar differential rotation regarding solar activity. Here we analyze solar cycles 12–24, investigating the long-term temporal variations in solar differential rotation. It is shown that $J_2$ exhibits a net modulation over the 13 studied cycles of $\approx (89.6 \pm 0.1)$ yr, with a peak-to-peak amplitude of $\approx 0.1 \times 10^{-7}$ for a reference value of $2.07 \times 10^{-7}$). Moreover, $J_2$ exhibits a positive linear trend in the period of minima solar activity (sunspot number up to around 40) and a marked declining trend in the period of maxima (sunspot number above 50). In absolute magnitude, the mean value of $J_2$ is more significant during periods of minimum than in periods of maximum. These findings are based on observational results that are not free of errors and can be refined further by considering torsional oscillations for example. They are comforted by identifying a periodic variation of the $J_2$ term evidenced through the analysis of the perihelion precession of planetary orbits either deduced from ephemerides or computed in the solar equatorial coordinate system instead of the ecliptic coordinate one usually used.

*Unified Astronomy Thesaurus concepts:* Solar physics (1476); Solar activity (1475); Solar rotation (1524); Sunspots (1653); Gravitation (661); Fundamental parameters of stars (555); Equatorial coordinate system (467); Ecliptic coordinate system (445); Solar evolution (1492); The Sun (1693); Solar motion (1507)

## 1. Introduction

It has been suggested by Dicke (1976), an astronomer at Princeton (USA), as early as the 1970s that the measured excess of solar oblateness over the oblateness due to the surface rotation alone might be due to the existence of a solar gravitational moment that in turn, could be due to a rapidly rotating solar core. Note that this thesis has been (re)brought up to date by Fossat et al. (2017) without any consideration, which could be drawn so far on the surface oblateness.

Let us recall that, in a spherical harmonics expansion in $n,m$ ($n$, order; $m$, mode) of the gravity potential outside a star, the gravitational moments are determined by the tesseral coefficients $c_{nm}$ falling off inversely as the cube of the distance from the star's center. Because the Sun is essentially symmetric about its rotation axis $m=0$; thus the second-order $n=2$, or zonal coefficient $c_{2,0}$, determines the quadrupole moment. As $c_{2,0}$ is always negative, by convention and simplification, $J_2$ is taken to be $-c_{2,0}$ (note that $J_2$ is the dynamical flattening and not the solar oblateness as it is sometimes written; see also footnote 1 in Pireaux & Rozelot 2003).

For a very long time, $J_2$ has hardly attracted interest due to two significant facts: on the one hand, its order of magnitude is very faint, and on the other hand, it cannot be measured directly; models are required. On the first point, Pireaux & Rozelot (2003) assigned $J_2$ to be $\approx (2.0 \pm 0.4) \times 10^{-7}$, a range of values now commonly accepted, sometimes slightly revised as $\approx (2.2 \pm 0.4) \times 10^{-7}$. On the second point, several indirect observations have been proposed. Among them, let us quote Armstrong & Kuhn (1999), who explored rotation models that smoothly match the observed surface rotation and interior measurements deduced from the helioseismic interior rotation. Other different methods of theoretical calculations have been advanced; one is to express the distortions of the solar shape under the assumption of a slow rotation (i.e., when the centrifugal acceleration is slight compared to the gravitational acceleration) and where all solar structure quantities are described in terms of perturbations (expanded based on Legendre polynomials) of the spherically symmetric nonrotating star. The gravitational moments $J_{2n}$ are thus obtained assuming the continuity of the gravitational potential at the solar surface (see, i.e., Equation (3) in Lefebvre et al. 2007 or Equation (17) in Mecheri et al. 2004). In this formulation, the $J_2$ determination represents solely the purely gravitational contribution, which is likely not entirely correct. Another less common method is considering a rotating star's equilibrium formation. Under the assumption of hydrostatic equilibrium, the body's shape is spheroidal in response to the self-gravitational and centrifugal potentials. Hence the shape is defined by the angular spin velocity and the radial density profile. Thermodynamical parameters render the analytical treatments complicated but possible when ellipticities are close to zero (generally associated with states of low rotation as in the solar case).

An alternative indirect approach is to access $J_2$ by analyzing the orbits of planets and asteroids of the solar system. For many years, the accuracy of the ephemeris of such bodies has been incredibly improved, and numerical solutions lead to determining $J_2$ by a postfit residual minimization. However, it is not simple to reach this goal because of the interplay between the effects of the solar multipolar moments with those induced by the post-Newtonian gravito-electromagnetic forces (Lense–Thirring effect; see Iorio 2018, and Section 5).

We are here interested in the $J_2$ variation with time, a dependence that has been hardly studied until now. We comprehend that our analysis is based on observations, which are subject to errors; the results will inevitably be affected. However, they do provide indications of the long-term behavior of $J_2$.

## 2. Evolution of $J_2$ over the Solar Cycles 12–24

Today the question of the temporal dependence of the gravitational moments $J_n$ (and so, $J_2$ at first) is not determined as (i) observations are at the cutting edge of the techniques and (ii) the mapping of the surface magnetic fields, which could produce a supplementary shape distortion (or not) due to the rotation, is not known with sufficient accuracy to be accurately modeled. The same approach goes for other factors that may be sensitive, such as turbulent pressure, shear effects, or other stresses, and would contribute to affecting the solar shape. However, contemporary measurements of the solar figure made utilizing the MDI-SOHO experiment (Scherrer et al. 1995) or by the Helioseismic and Magnetic Imager instrument on board the Solar Dynamics Observatory (Scherrer et al. 2012) indicate a temporal variability of the asphericities' coefficients (see, i.e., Emilio et al. 2007; Kuhn et al. 2012, and Kosovichev & Rozelot 2018). Even if the contribution of the solar limb shape through these parameters is a few percent due to the gravitational moments, it turns out that temporal variability is expected. Indeed, determining their order of magnitude this way requires very high sensitivity methods.

Helioseismology provided the premise for a variation of the gravitational moments associated with the solar cycle. This has been highlighted by Antia et al. (2008), who found an amplitude modulation of less than 0.04% for $J_2$ over the time range 1996–2006. However, such a tiny modulation has not been confirmed so far.

$J_2$ was here computed as usual by setting the gravity field. As we wanted only to highlight the temporal dependence, we determined $J_2$ at the latitude at which the rotational gradient $(\partial \log(\omega)/\partial \log(r))$ is reversing, passing from negative to positive values. Indeed, the (logarithmic) average gradient in the outer 15 Mm or so is close to $-1$ and is quite independent of latitude below 30°; between 30° and 50° latitude, it is still negative but makes a transition to absolute value at 56° of latitude (Corbard & Thompson 2002).[3] At this specific latitude, the centrifugal force that affects the solar shape can be derived from the potential as the observed surface rotation is very similar to the equatorial excess. This simplifies the calculations without loss of generality, bearing in mind that the geodetic parameter $q = (\omega^2 R^3)/GM$ remains a small quantity, albeit latitude dependent. Taking $M_\odot = 1.989 10^{30}$ kg (solar mass), $G = 6.6726 10^{-11}$ m$^3$ kg$^{-1}$ s$^{-2}$ (gravitational constant), and adopting the following values for the solar parameters: $R_{eq} = 695\ 509.9835$ km (solar equatorial radius), $R_{pol} = 695\ 504.0331$ km (polar radius) in such a way that $R_\odot = 695\ 508.0000$ km (astrometric accuracy but needed)[4], and $f = 8.56 \times 10^{-6}$, it leads to $J_2 = +2.07 \times 10^{-7}$ at 56° of latitude, considering the commonly adopted rotation law $\omega(\theta) = A + B \sin^2(\theta)$, where $A$ and $B$ are explained hereafter. The above-deduced value of $J_2$ adopting Javaraiah's data compiled in Javaraiah (2020) is labeled in the remainder of this article as the "reference" value $J_{2\mathrm{ref}}$.

It was noticed in the 1980, first by Lustig (1983), then by Balthasar et al. (1986), from the observations of the surface that the differential rotation profile (given by the formula mentioned above $\omega(\theta)$) exhibits variations with the solar cycle. Many other authors, such as Howard et al. (1980) and Li et al. (2014) reanalyzed the observational data of the differential rotation law (including the subsurface interior). The results did not seem convincing enough, until the 2000s, due to a too short sample length. Using the Greenwich sunspot-group data compiled for 1874–2017, Javaraiah (2020) investigated the long-term variations in solar differential rotation versus sunspot activity. He determined the equatorial rotation $A$ and the latitude gradient $B$ components around the maxima and minima of solar cycles 12 to 24. He found a significant temporal dependence between these two components.

We used data listed in Javaraiah (2020) Table 2, giving values of $A$ and $B$ (in degree per day), in the 3 yr intervals at the epochs Tm and TM (middle years of the corresponding 3 yr intervals) of the minima and maxima (indicated by the suffixes m and M, respectively) of solar cycles 12–24. The minimum (rm) and maximum (RM) sunspot numbers are also given in the quoted Table. We added, for a more detailed analysis, the values of the sunspot number taken from WDC-SILSO data (see SILSO 2022) at the mean epoch of each solar cycle. The uncertainties have been computed from the uncertainties given by Javaraiah (2020) on $A$ and $B$.

## 3. Analysis of the Results

Computations of $J_2$ using the abovementioned data sets enable us to plot (the following figures): $J_2$ as a function of date, from 1872 to 2010 (solar cycles 12 to 24), $J_2$ again as a function of date, but at the epochs around the maxima (TM) and minima (Tm) of the studied solar cycles. As shown in Figure 1, $J_2$ is the function of the sunspot number (SILSO values), around the minima, then around the maxima, and for the whole data set. The following sections are present the analysis in this order: according to the date (Section 3.1), the maximum and minimum periods ranging over time (Section 3.2), and then the whole data set according to the global solar activity (Section 3.3). Conclusions and perspectives are drawn in Section 5.

### 3.1. Analysis versus the Considered Span Time 1872 to 2010

In Figure 1, we present the results for $J_{2\,(TM)}$ (top left), $J_{2\,(Tm)}$ (top right), and $J_{2\,(whole)}$ (bottom) versus the date. We first note that the means and their standard variations are, respectively, $(2.00 \pm 0.01) \times 10^{-7}$, $(2.06 \pm 0.04) \times 10^{-7}$, and $(2.01 \pm 0.02) \times 10^{-7}$. The $J_{2\,(Tm)}$ estimate is thus of the same order as the reference value ($J_{2\mathrm{ref}} = 2.07 \times 10^{-7}$, see Section 2), while the two others are slightly lower, suggesting that the mean magnitude of $J_2$ is higher in periods of reduced solar activity (and vice versa, i.e., less in periods of higher solar activity). Taking, as usual, a significant probability level $p$ at 0.05, a mere estimate of the three different $J_2$ tested against the expected reference value through a student's test, gives the following confidence $s$ intervals (in $10^{-7}$): [2.02–2.10] for ($J_{2\,\min}$), [1.98–2.07] for $J_{2\,\max}$, and [2.00–2.07] for $J_{2\,\mathrm{(whole)}}$. Thus, $J_{2\,\min}$ is significantly of $A$ and $B$ from the observations ($p = 0.07$); we find the following probabilities for $J_2$ to be in the range of $[1.95–2.1] \times 10^{-7}$: 0.71 ($J_{2\,\mathrm{max}}$), 0.43 ($J_{2\,\mathrm{min}}$), and 0.93 ($J_{2\mathrm{whole}}$). The probability rises to 0.71 for $J_{2\,\mathrm{min}}$ to be in the range $[1.95–2.30] \times 10^{-7}$. These last findings give some confidence in the results.

---
[3] There is a region of a positive gradient in the outer 5 Mm at high latitudes, a depth that has also been evidenced by Kosovichev & Rozelot (2018).

[4] $R_{eq}$ and $R_{pol}$ are unequivocally determined since $R_\odot$ and $f$ are, as $R_\odot = [R_{eq}^2 \times R_{pol}]^{1/3}$ and $f = R_{eq}/R_{pol}$.

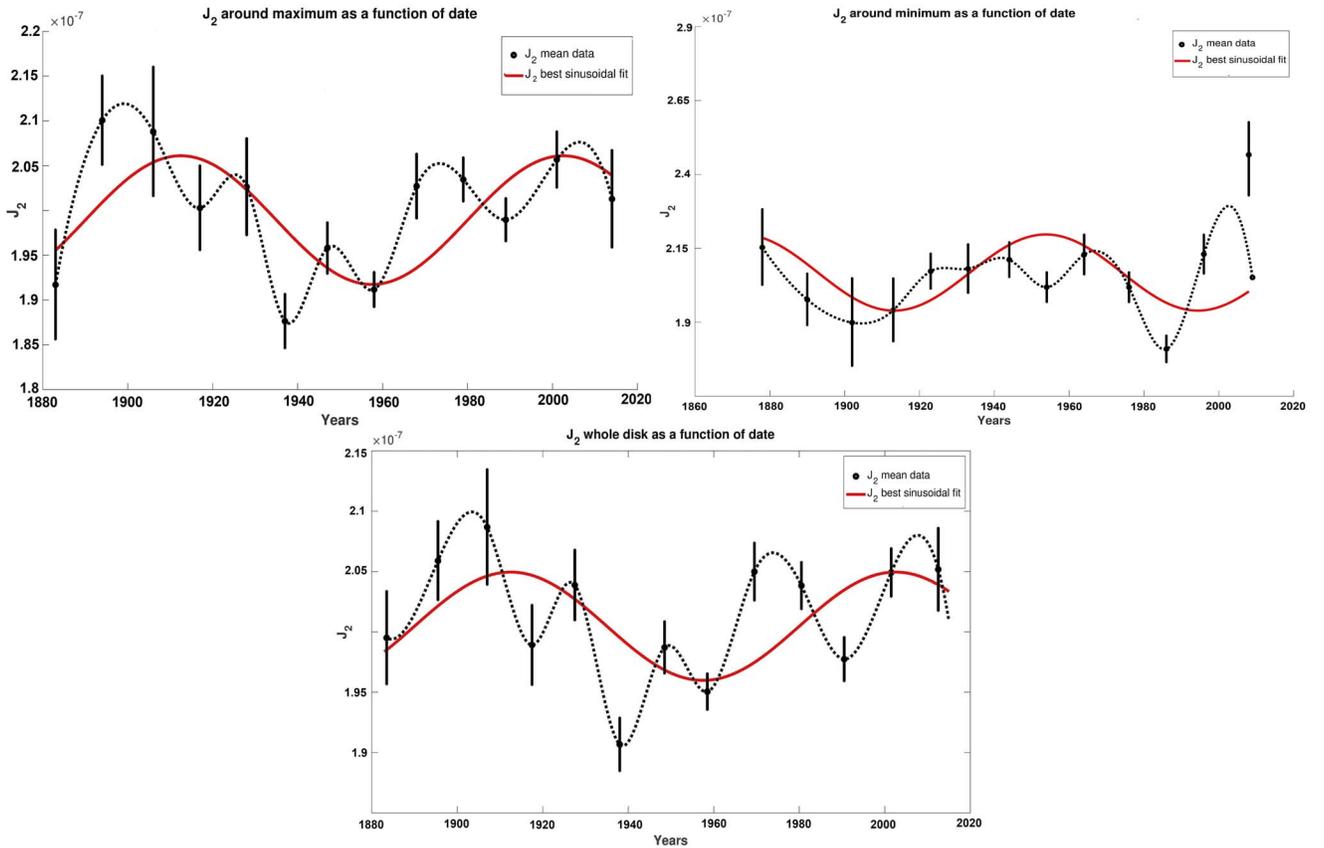

**Figure 1.** Solar quadrupole moment $J_2$ as a function of time (1872 to 2010), considering the mean epoch of each solar cycle 12 to 24 (bottom). Top left, at the epochs around the maximum of solar activity TM, then, top right, at the epochs around the minimum Tm. A clear periodic oscillation is visible on all the data, evidencing a temporal dependence of the first gravitational moment, of around $(89.6 \pm 0.1)$ yr. For the sake of clarity, error bars have been multiplied by 20.

The inspection of the three plots in Figure 1 (TM, Tm, and the whole data) according to the date reveals a net sinusoidal oscillation whose periods are, respectively, $P_{\mathrm{RM}} = (89.5 \pm 0.2)$ yr, $P_{\mathrm{rm}} = (80.3 \pm 1.5$ yr), and $(P_{\mathrm{whole}} = 89.6 \pm 0.1)$ yr, with significant correlation coefficients: $r = 0.45$, 0.70, and 0.61. The weighted average (by errors) of the found period is thus $(89.56 \pm 0.09)$ yr. The largest amplitude of the modulation is $\approx 0.1 \times 10^{-7}$ (peak to peak). The uncertainties on the amplitude of the sinusoidal fits are, respectively, about $2.0 \times 10^{-8}$, $2.3 \times 10^{-8}$, and $2.1 \times 10^{-8}$.

An outlier can be noticed for cycle 24 ($2.47 \times 10^{-7}$) due to the value of $A$ in Javaraiah (2020)'s Table 2:14.384 (degree per day), which is not logical. If we remove this point, the correlation coefficient becomes 0.65 (without affecting the period). However, we keep it, and this remark on this outlier point stands for the remainder of this paper.

Over the studied temporal span (solar cycles 12 to 24), the three detected periodicities show that the oscillation would be somewhat shorter during periods of solar minimum. If such an issue proves to be accurate, we will try to explain it in Section 5, but suggesting (if the values found are significantly different) that $J_2$ could be more time sensitive with periods of minimum solar activity. The following sections will attempt to check such findings.

As a partial conclusion, considering the errors in the experimental parameters, we may infer a periodic oscillation of the first gravitational moment of about eight solar cycles, which roughly corresponds to the so-called Gleissberg cycle (Gleissberg 1939), generally taken to be equal to $\approx 87$ yr (or eight solar cycles).

### 3.2. Analysis versus the Periods around the Solar Maxima and Minima

Figure 2 depicts the $J_2$ behavior with a date of around the 3 yr, enclosing the maximum (left) and the minimum (right). A very slight declining trend can be noticed also in the second case that we did not plot as almost identical to the mean. Here, $J_{2\,\mathrm{min}}$ does not significantly differ from the mean. The presence of the outlier does not significantly influence this result; when removed, a slightly positive trend emerges.

The situation is different around the maximum: a net negative trend ($r = -0.46$) for a mean estimated at $2.00 \times 10^{-7}$. In any case, the $J_2$ magnitude seems more significant in the period of minimum activity than in the period of maximum, a situation that we will find again by studying the $J_2$'s dependence on the sunspot number (Section 3.3).

To partially conclude, this analysis shows that $J_2$ behaves differently during the various phases of the solar cycle.

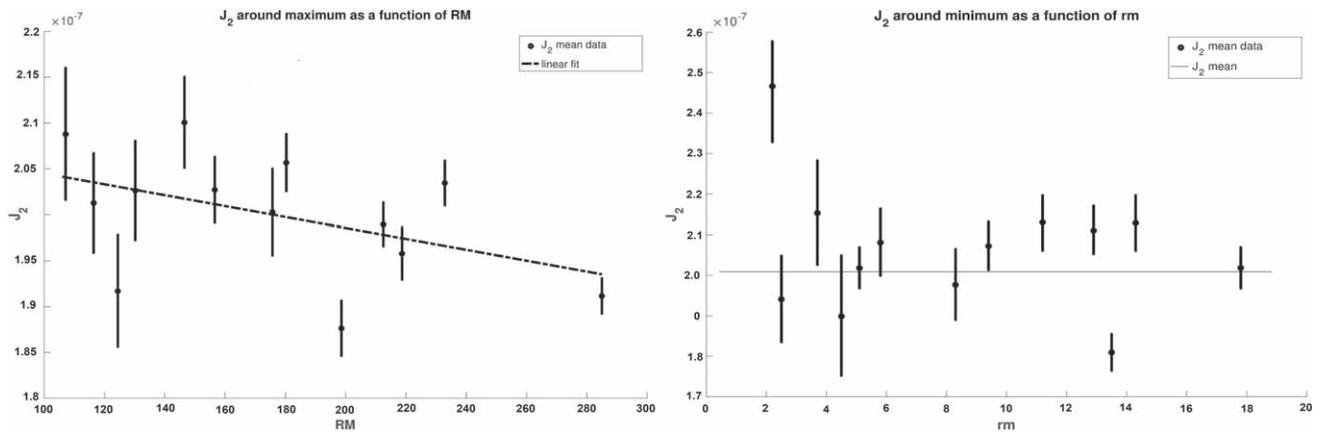

**Figure 2.** Solar quadrupole moment $J_2$ vs. the solar activity analyzed during the 3 yr span time around the periods of maximum (RM; left) and minimum (rm; right). During periods of solar minima, $J_2$ does not significantly differ from the mean ($2.06 \times 10^{-7}$, thin line), which is of the same order of the $J_2$ reference value. During periods of solar maxima, $J_2$ shows a marked negative trend ($r = 0.46$). For the sake of clarity, error bars have been multiplied by 20.

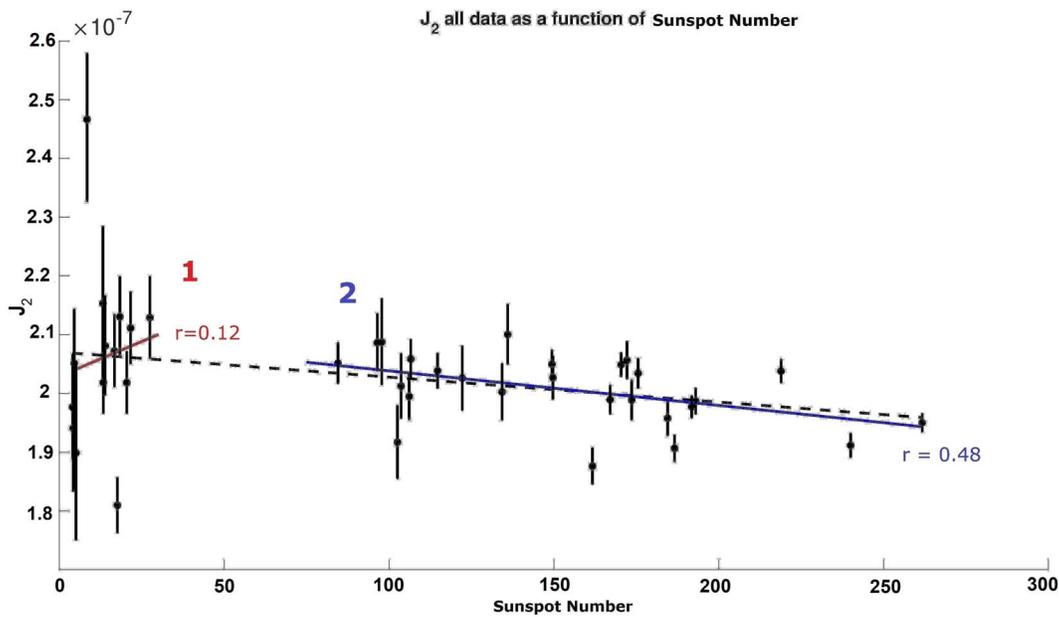

**Figure 3.** Solar quadrupole moment $J_2$ as a function of the solar activity (1872 to 2010) for the whole data available. Although the general trend seems to be negative (dotted line), the first solar quadrupole moment seems to follow two regimes: during periods or minimum activity (sunspots number 0–40, left leg), $J_2$ is positively correlated, while during periods of maximum of activity (80–200, right leg), the trend is negative. For the sake of clarity, error bars have been multiplied by 20.

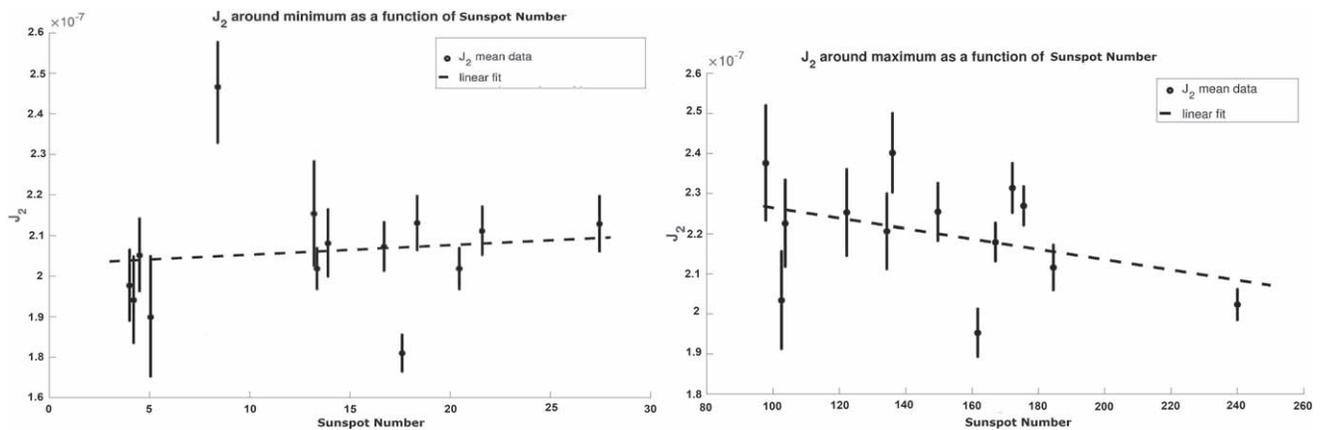

**Figure 4.** Solar quadrupole moment $J_2$ as a function of the solar activity (1872 to 2010), described by the sunspot numbers. Top left, for small sunspot numbers (0–40), the trend is slightly positive: $r = +0.12$. Top right, for higher sunspot numbers (80–200), the trend is clearly negative: $r = -0.62$. For the sake of clarity, error bars have been multiplied by 20.

### 3.3. Analysis versus the Solar Activity Described by the Sunspot Numbers

In Figure 3, we present the results for all the data available according to the sunspot number. The $J_2$ temporal evolution depicts a somewhat more complex behavior. Although the overall trend is negative when plotting the whole data as a function of the sunspot number, it appears a positive trend for small values (0–40) (marked (1)) and a negative trend for larger values (80–250) (marked (2)). Splitting the data into these two series yields Figure 4 showing how the solar quadrupole moment $J_2$ evolves as a function of the solar activity (1872 to 2010). Left, for small values of the sunspot numbers, during the 3 yr of the minimum (rm: 0–40), the trend is slightly positive ($r = +0.12$). Right, for higher sunspot numbers (RM: 80–200), around the maximum 3 yr span, the trend is negative ($r = -0.48$).

If the obvious outlier is removed in Figure 4 (left, i.e., $2.47 \times 10^{-7}$), the correlation coefficient jumps to $r = +0.69$. Thus, the identification of the two regimes is well highlighted.

Together with the 89 yr period, a shorter one (a subharmonic) is detected in the three plots shown in Figure 1, respectively, (max-min-whole): $8.80 \pm 0.20$, $8.53 \pm 0.11$, and $9.20 \pm 0.10$ yr, with correlation coefficients of 0.67, 0.51, and 0.44. The weighted average is thus $8.89 \pm 0.07$ yr. (Note that the periodicity found during the period of minimum is less than the two others.) This issue could be related to the oscillations of short timescales ranging from $\approx 0.5$ yr to $\approx 9$ yr (short-term variations of $\approx 155$ days are called Rieger periodicities, more frequently found in epochs of maximum activity). The longer periodicities are related to the quasi-biennial oscillation (QBO), whose lengths are poorly defined while their amplitudes are modulated by the solar cycle. Their signatures have been seen in solar rotational rate residuals at near-surface depths (Inceoglu et al. 2022), and it has been shown that their relative amplitude is highly correlated with sunspot number.

They also suggest that the amplitude of the QBO in both frequency shift data and solar activity proxy data scales with the activity of the solar cycle. We drew such an inference in our results. Explanations have been put forward; among them, Jupiter, Saturn, Uranus, and Neptune) could influence the solar differential rotation rate and hence, the strength of the solar dynamo (Javaraiah 2020; de Paula et al. 2022; Zioutas et al. 2022). We are wondering if our findings could not highlight any relation of $J_2$ with such issues, as well as on long timescales of the solar cycles, and even in shorter ones, as detected, for instance, in the variability of the sunspots N–S asymmetry activity (see for instance de Paula et al. 2022): "the 7.0-07.9 yr periodicity could not be an artifact but a real signal in the solar N–S asymmetry and only its fluctuating apparition along the time can explain why it evaded its manifestation in the analysis of other preceding works. Future solar models should integrate this period."

## 4. Discussion

Is the differential rotation law on which this study is based sufficiently valid? As sunspots only cover a rather small range of latitudes, it can be argued that the overall surface differential rotation is better described by a law of type $\omega(\theta) = A + B \sin^2(\theta) + C \sin^4(\theta)$. Based on magnetogram data obtained at Mount Wilson (USA), Howard et al. (1980) determined these coefficients $A$, $B$, and $C$. They suggested that a strong correlation between $B$ and $C$ might occur, albeit some authors consider it spurious. They also showed that a simple linear relationship between three coefficients can be constructed (orthogonal functions), eliminating the crosstalk among the coefficients and for rotation, providing a convenient set of functions (Gegenbauer polynomials) for separating modes in torsional oscillations. At first sight, Howard's law seems preferable. However, the Mount Wilson survey was performed only over the years 1973–1977, i.e., over only 4 yr, which seems insufficient for our purpose. A reanalysis of the data in 1984 (Howard et al. 1984) covered the years 1921–1982, which is 61 yr and would be better. But they determine a law in only $A + B \sin^2(\theta)$, which however fits well the observations. Snodgrass & Howard (1985) established again a law in $A + B \sin^2(\theta) + C \sin^4(\theta)$, using Gegenbauer polynomials, for the years 1967–984, i.e., 17 yr covering hardly two solar cycles, SC 20 and SC 21. Sometime later, Snodgrass & Ulrich (1990) reexamined this law over 20 yr (from 1967 to 1987), also considering the Gegenbauer polynomials. Thus, this analysis covers approximately the two same solar cycles (SC 20 and SC 21), which is again a bit limited for our purpose. All these reasons led us to a first approach to consider only Javaraiah's law (2020), essentially also because the coefficients $A$ and $B$, which are time-dependent, are tabulated over 13 solar cycles, i.e., 137 yr, which moreover makes the detected period significant.

However, it seemed interesting to compare $J_2$ as deduced from Howard's law (1980) considering $A$, $B$, and $C$, which gives $2.18 \times 10^{-7}$. This is not fundamentally different from our reference value $2.07 \times 10^{-7}$, and is in the error limits $(2.1 \pm 0.4) \times 10^{-7}$ (Pireaux & Rozelot 2003). Note that the results obtained by Snodgrass & Ulrich (1990), determined by Doppler features, are about 4% higher than those deduced from Mount Wilson spectroscopic observations, and would lead to a higher $J_2$ estimate ($>2.88 \times 10^{-7}$). It can be assumed that the visualization of the curves obtained by this new method, at least to first order, would only differ from those obtained in this paper by a simple translation in ordinates.

Regarding torsional oscillations, as seen just before, an $A + B \sin^2(\theta)$ law does not capture them very well. These oscillations, which periodically speed up or slow down the rotation in certain zones of latitude, mainly accentuated at high latitudes ($>62°$, while elsewhere the rotation remains essentially steady), could probably be considered later because modern methods covering the whole solar disk are not yet available for the timescale of the Gleissberg cycle. For the time being, such an analysis is clearly beyond the scope of this study, for which we only wanted to show that $J_2$ might be temporally dependent (or not).

Figure 1 shows a different behavior at minimum (or more precisely around the minimum, and not necessarily during the whole minimum) and at maximum (or more precisely around the maximum, and not necessarily during the whole maximum). Such a complex behavior has already been detected by Emilio et al. (2007) from MDI measurements on SOHO between 1997 (period of minimum) and 2001 (period of maximum), suggesting that the outer solar atmosphere expands nonhomologously during the cycle. This result was also found by Rozelot et al. (2009a) who moreover showed that there could be a change in the relative importance of the hexadecapolar term and the dipolar one in the course of the activity cycle. In times of high activity, only the first moment has a significant contribution, but in times of low activity, the second one is predominant. This could be interpreted also as a periodic angular momentum exchange between the photosphere and deeper layers of the convection zone. Such a complex mechanism takes certainly its root in the leptocline (for the definition see footnote 6 in the abovementioned paper and Figure 6 in Rozelot et al. 2009b); a zone in which it has been shown that the rotational gradient is nonconstant, which might explain also our results.

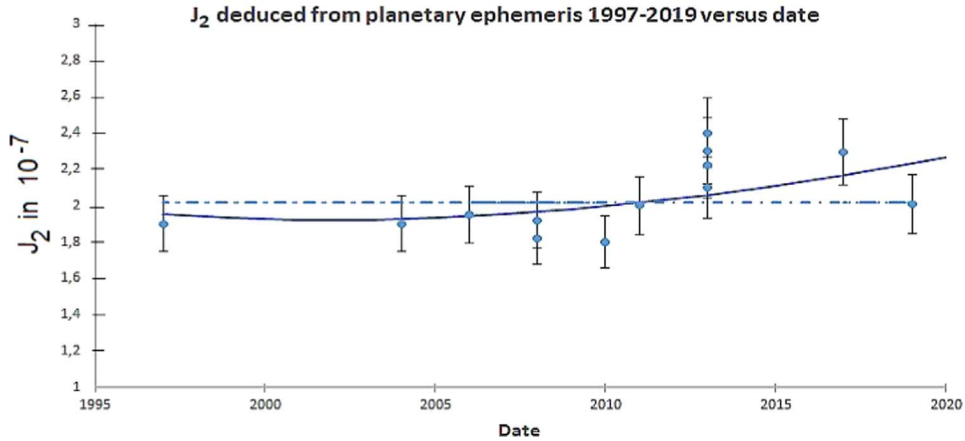

**Figure 5.** Solar quadrupole moment $J_2$ deduced from solutions to the planetary motion fitted to observational data permitting to assign estimates to all unconstrained ephemeris parameters. The solid line represents a part of a long period signal of 88 yr ($J_2 = 2.04 \times 10^{-7} + 3.00 \times 10^{-9} \sin(2\pi \times \text{date}/88)$; $r = 0.8$. For the sake of clarity, the sine function ordinate was multiplied by (16). The dotted line represents the mean: $2.02 \times 10^{-7}$.

Another point is raised about possible changes in the internal rotation model that do not necessarily appear in the surface rotation data. Recent 3D simulations (Kitiashvili et al. 2022) supported by observations have shown that the rotational effects in solar subsurface convection produce the formation of rotational shear and meridional circulation at midlatitudes. The structure of this near subsurface layer is "not uniform but contains a sharp shear layer in the top ≈8 Mm," which has been identified as a leptocline (see also Li et al. 2022) in which radial variations of the differential rotation occurs, contributing obviously to more complex rotational laws.

All these issues will be addressed in later works.

## 5. Conclusion

The long-term variations in solar differential rotation revealed that the first gravitational moment $J_2$ seen from the surface distortion is variable. The current temporal evolution for the last 138 yr (2010–1872) shows a periodic modulation of about (89.6 ± 0.1) yr, of ≈0.1 × $10^{-7}$ modulation amplitude.

If such a period of ≈8 solar cycles is highlighted, it could then be associated with the Gleissberg period.

We have identified that $J_2$ seems to follow two regimes. In the period of minimum solar activity, the mean magnitude order of $J_2$ is around the reference mean and shows a positive trend with increasing sunspot numbers from 0 up to around 30–40. By contrast, in the period of maximum solar activity, the mean magnitude order of $J_2$ is less than the reference mean and shows a declining trend with increasing sunspot numbers from around 40 (up to 200 and more).

We want to emphasize the importance of determining $J_2$ with great precision, as this parameter also plays a vital role in relativistic astrometry and relativistic celestial mechanics (Rozelot & Fazel 2013). In this last paper, $J_2$ was compiled from several observations since 1877, with modern observations starting from 1966. It already inferred a dynamical flattening, seeming unlikely at that time but a bit more realistic in light of this study.

Let us briefly comment on the two fields mentioned above while avoiding confusion between the potential temporal variation of the solar shape and estimating it together with Gravitational Rotation (GR) parameters.

First, regarding the parameterized post-Newtonian (PPN) modern theories, it has been shown that the PN parameters $\beta$ (which encodes the amount of nonlinearity in the superposition law of gravitation) and $\gamma$ (which encodes the amount of curvature of spacetime per unit rest-mass) are linked to the solar quadrupole moment $J_2$ through a linear relation. Even though it would be possible to extract $J_2$ from planetary ephemerides in principle, it is significantly correlated with other solution parameters (semimajor axis of planets, the mass of asteroids...). Focusing on the $J_2$ correlations, Rozelot et al. (2022) have found that, in general, the correlations $[\beta, J_2]$ and $[\gamma, J_2]$ are ≈45% and ≈55%, respectively. In this respect, the contribution of the quadrupole competes with that of PN parameters of the order of $10^{-2}$. The effect of a change in $\beta$ can be distinguished from a change in $\gamma$; a determination of other significant figures of $J_2$ is equally essential to be able to say anything significant about the PN parameters (Sebastián et al. 2022).

The situation could be improved with additional spacecraft measurements, but it remains challenging. We are still waiting for results from space missions for which precision at a level of $10^{-8}$ (or more) is expected; thus, it can be awaited to highlight a $J_2$ temporal dependence. In this context, in exploring the available $J_2$ values deduced from the precession of Mercury's perihelion along the orbit plane due to the Sun's quadrupole moment, we have shown its possible variation with time (Rozelot & Eren 2020).

Figure 5 shows the solar quadrupole moment $J_2$ deduced from solutions to the planetary motions (especially Mercury), fitted to observational data retrieved from 14 contributions ranging from 1997 to 2019. This figure is extracted from the data used in Rozelot & Eren (2020), Table 1, for which we added two measurements since new values were available in 2017 and 2019 (see data in "Notes Scientifiques et Techniques de l'Institut de Mécanique Céleste", 2017, 2019). The computation process permits assigning estimates to all unconstrained ephemeris parameters so that $J_2$ can finally be obtained. It should be noted that even if the sample used, of no more than 12 yr, is much smaller than the one used in this study (137 yr), it appears to have modulation of $\approx 88$ yr. Note that the sine function $J_2 = 2.04 \times 10^{-7} + 3.00 \times 10^{-9} \sin(2\pi \times \text{date}/88)$ gives a modulation of the amplitude of $0.06 \times 10^{-7}$ (3%) fully compatible with GR.

Regarding the second item, the secular solutions for the oblateness disturbance in consideration of the periodic variation of the $J_2$ term have been studied by Xu et al. (2017) to derive the perihelion precession of Mercury. The results show that the difference in Mercury's perihelion precession between the solar equatorial plane and the ecliptic plane can reach a magnitude of $126\,708 \times J_2$, which is even more noteworthy than the perihelion precession itself ($101\,516 \times J_2$). In this context, when a periodic variation of the $J_2$ term is considered, instead of simply a constant, the periodic $J_2$ has an effect of nearly 0.8% of the secular perihelion precession of Mercury. This indicates that a better understanding of solar oblateness is required, which could be done, for instance, through observation in the solar orbits instead of on Earth.

Finally, we would like to point out that the Pioneer's anomalous acceleration could be explained utilizing the observed solar quadrupole moment. Indeed, it generates an acceleration of the same order of magnitude as Pioneer's constant acceleration, within the accuracy range of the observed anomalous acceleration (Quevedo 2005). Hence the need for greater precision on $J_2$.

In a general conclusion, we have underlined the importance of knowing the temporal variations of the first solar gravitational moment $J_2$ with remarkable accuracy and its changing behavior with the solar cycle that may lead to a better understanding of the physical phenomena involved in the leptocline. It shows that a long periodic oscillation of $J_2$ of the order of the Gleissberg's period, which has never been put in evidence before, with amplitude modulation of the order of less than 5%. Considering the errors inevitably linked to the observations (we detected an outlier in Javaraiah's data, but we voluntarily kept it), the true modulation is certainly lower. It is, however, fully compatible with the GR. The impact of $J_2$ is challenging to better determine its role in solar activity, suggesting some form of correlation between QBOs and even N–S solar asymmetry variability.

The authors thank the referee for valuable remarks made on the article, which led to better discussion of the results.

## ORCID iDs

Saliha Eren  https://orcid.org/0000-0001-7603-2488
Jean-Pierre Rozelot  https://orcid.org/0000-0002-5369-1381

## References

Antia, H. M., Chitre, S. M., & Gough, D. O. 2008, A&A, 477, 657
Armstrong, J., & Kuhn, J. R. 1999, ApJ, 525, 533
Balthasar, H., Vázquez, M., & Wöhl, H. 1986, A&A, 155, 87
Corbard, T., & Thompson, M. J. 2002, SoPh, 205, 211
de Paula, V., Curto, J. J., & Oliver, R. 2022, MNRAS, 512, 5726
Dicke, R. H. 1976, SoPh, 47, 475
Emilio, M., Bush, R. I., Kuhn, J., & Scherrer, P. 2007, ApJ, 660, L161
Fossat, E., Boumier, P., Corbard, T., Provost, J., Salabert, D., et al. 2017, A&A, 604, A40
Gleissberg, W. 1939, Obs, 62, 158
Howard, R., Boyden, J. E., & Labonte, B. J. 1980, SoPh, 66, 167
Howard, R., Gilman, P. I., & Gilman, P. A. 1984, ApJ, 283, 373
Inceoglu, F., Howe, R., & Loto'aniu, P. T. 2022, ApJ, 925, 170
Iorio, L. 2018, in The Fourteenth Marcel Grossmann Meeting On Recent Developments in Theoretical and Experimental General Relativity, Astrophysics, and Relativistic Field Theories: Proc. of the MG14 Meeting on General Relativity, ed. M. Bianchi, R. T. Jantzen, & R. Ruffini (Singapore: World Scientific), 3679
Javaraiah, J. 2020, SoPh, 295, 1
Kitiashvili, I. N., Kosovichev, A., Wray, A., Sadykov, V., & Guerrero, G. 2022, MNRAS, 518, 504
Kosovichev, A., & Rozelot, J.-P. 2018, ApJ, 861, 90
Kuhn, J. R., Bush, R., Emilio, M., & Scholl, I. F. 2012, Sci, 337, 1638
Lefebvre, S., Rozelot, J. P., & Kosovichev, A. G. 2007, AdSpR, 40, 1000
Li, K. J., Feng, W., Shi, X. J., Xie, J. L., Gao, P. X., et al. 2014, SoPh, 289, 759
Li, K., Xu, J., & Feng, W. 2022, NatSR, 12, 1
Lustig, G. 1983, A&A, 125, 355
Mecheri, R., Abdelatif, T., Irbah, A., Provost, J., & Berthomieu, G. 2004, SoPh, 222, 191
Pireaux, S., & Rozelot, J.-P. 2003, Ap&SS, 284, 1159
Quevedo, H. 2005, in AIP Conf. Proc. 758, Gravitation and Cosmology: 2nd Mexican Meeting on Mathematical and Experimental Physics (Melville, NY: AIP), 129
Rozelot, J., Damiani, C., & Pireaux, S. 2009a, ApJ, 703, 1791
Rozelot, J.-P., Damiani, C., & Lefebvre, S. 2009b, JASTP, 71, 1683
Rozelot, J. P., & Eren, S. 2020, AdSpR, 65, 2821
Rozelot, J., & Fazel, Z. 2013, Solar Dynamics and Magnetism from the Interior to the Atmosphere (Berlin: Springer), 161
Rozelot, J.P., Kilcik, A., & Fazel, Z. 2022, arXiv:2208.06779
Scherrer, P. H., Bogart, R. S., Bush, R., et al. 1995, The SOHO Mission (Berlin: Springer), 129
Scherrer, P. H., Schou, J., Bush, R. I., et al. 2012, SoPh, 275, 207
Sebastián, A., Acedo, L., & Moraño, J. 2022, AdSpR, 70, 842
SILSO 2022, International Sunspot Number Monthly Bulletin and online catalogue, https://www.sidc.be/silso/
Snodgrass, H. B., & Howard, R. 1985, SoPh, 95, 221
Snodgrass, H. B., & Ulrich, R. K. 1990, ApJ, 351, 309
Xu, Y., Shen, Y., Xu, G., Shan, X., & Rozelot, J.-P. 2017, MNRAS, 472, 2686
Zioutas, K., Maroudas, M., & Kosovichev, A. 2022, Symm, 14, 325